\documentclass[prb,twocolumn,aps]{revtex4}
\usepackage[latin1]{inputenc}
\usepackage{latexsym,amstext,amssymb,amsmath}
\usepackage[dvips]{graphicx}
\usepackage{color}
\begin{document}
%%%%%%%%%%%%%%%%

%%%%%%%%%%%%%%%%%%%%%%%%%%%%%%%%%%%%%%%%%%%%%%%%%%%%%%%%%%%%%%%%%
\title{Lattice dynamics and structural stability of ordered
  Fe$_3$Ni, Fe$_3$Pd and Fe$_3$Pt alloys using density functional theory}
%%%%%%%%%%%%%%%%%%%%%%%%%%%%%%%%%%%%%%%%%%%%%%%%%%%%%%%%%%%%%%%%%     
 \author{M. E. Gruner$^{1}$}
 \email{Markus.Gruner@uni-due.de}
 \author{W. A. Adeagbo$^{2}$}
% \email{waheed.adeagbo@physik.uni-halle.de}
 \author{A. T. Zayak$^{3,4}$}
% \email{azayak@lbl.gov}
 \author{A. Hucht$^{1}$}
% \email{fred@thp.uni-duisburg.de}
 \author{P. Entel$^{1}$}
% \email{entel@thp.uni-duisburg.de}
 \affiliation{$^{1}$Faculty of Physics and Center for Nanointegration, CeNIDE, University of Duisburg-Essen, 47048 Duisburg, Germany}
 \affiliation{$^{2}$Institute of Physics, Martin Luther University Halle-Wittenberg, 06120 Halle, Germany}
 \affiliation{$^{3}$Department of Electrical Engineering and Computer
   Sciences, University of California, Berkeley, CA 94720, USA}
 \affiliation{$^{4}$Molecular Foundry, Lawrence Berkeley National Laboratory, Berkeley, CA 94720, USA}
\date{\today}
%%%%%%%%%%%%%%%%
\begin{abstract}
We investigate the binding surface along the Bain path and phonon
dispersion relations for the cubic phase of the ferromagnetic binary alloys
Fe$_3X$ ($X$$\,=\,$Ni, Pd, Pt) for L1$_2$ and D0$_{22}$ ordered phases 
from first principles by means of density functional theory.
The phonon dispersion relations exhibit a softening of the transverse
acoustic mode at the M-point in the L1$_2$-phase in accordance with
experiments for ordered Fe$_3$Pt. This instability can be associated
with a rotational movement of the Fe-atoms around the Ni-group element
in the neighboring layers and is accompanied by an extensive
reconstruction of the Fermi surface. In addition, we find an
incomplete softening in [111] direction which is strongest for Fe$_3$Ni.
We conclude that besides the valence electron density also the
specific Fe-content and the masses of the alloying partners should be
considered as parameters for the design of Fe-based functional
magentic materials. 
\end{abstract}
%%%%%%%%%%%%%%
\maketitle
%%%%%%%%%%%%%%%%%%%%%%
\section{Introduction}
%%%%%%%%%%%%%%%%%%%%%%
Ferrous alloys are well known for a wide variety of anomalous
structural and magnetic properties.
Around 1185\,K, bulk iron undergoes 
a structural transition from a face centered cubic (fcc)
high-temperature phase to a low-temperature
body centered cubic (bcc) phase, which is  ferromagentic
with a Curie temperature of 1043\,K.
The structural transition temperature is systematically lowered
by alloying elements from the right side of Fe
in the periodic table, which
effectively lowers the 
valence electron concentration $e/a$.
The transition is
displacive and diffusionless 
between a high-temperature high-symmetry austenitic phase and a
low-temperature martensitic phase, in which the symmetry is usually lowered,
e.\,g., by tetragonal distortions.
In ferromagnetic Fe-alloys with $e/a$ between 8.7 and 8.5, the martensitic
transition disappears and fcc austenite becomes the stable ground
state structure.
Iron alloys with Ni-group elements at this
valence electron concentration 
are at the center of scientific interest for a long time, as they
show, in addition to the martensitic instability, a multitude of
magneto-structural anomalies that are
of technological importance.
The probably most prominent is the so-called Invar effect, describing the
over-compensation of thermal expansion in a wide temperature range, 
which has been intensively studied for more than one century.\cite{cn:Guillaume97,cn:Wassermann90,cn:Shiga94}
The best known representative is Fe$_{65}$Ni$_{35}$, but similar
behavior is also observed
in Fe$_{70}$Pd$_{30}$ and Fe$_{75}$Pt$_{25}$. Although most
explanations proposed so far rely on magneto-volume coupling, the
vicinity to the martensitic transformation has been well noticed
and discussed. The relevant literature on this topic is too numerous
to be summarized at this point --
for an introduction and further references,
see, e.\,g., Refs.\ \onlinecite{cn:Nishiyama78,cn:Pepperhoff01,cn:Wassermann05}.

During the past three decades, the so-called
magnetic shape memory (MSM) effect, which is again present in the
alloys with all three Ni-group elements, was brought to the
attention of the scientific community.
The MSM behavior allows very large magnetic-field-induced
strains of up to three percent to be achieved by externally applying moderate
magnetic fields in the (sub-)Tesla range.\cite{cn:James98,cn:Hayashi00,cn:Kakeshita02,cn:Fukuda04,cn:Kakeshita06}
The effect is not as pronounced as for the Ni-based Heusler alloys
like Ni$_2$MnGa,\cite{cn:Ullakko96,cn:Sozinov02,cn:Soederberg05}
%in the martensitic state,
but its origin is believed to be related.
One common key ingredient is the appearance of intermediate or 
modulated martensites
-- 5-fold and 10-fold modulated pseudo-tetragonal or orthohombic in
the case of Ni-Mn-Ga\cite{Chernenko-98PRB,cn:Khovailo05,cn:Richard06}
and (slightly distorted) face centered tetragonal
(fct) in the case of Fe-Pt and Fe-Pd -- between the high symmetry
austenite and the low-temperature martensite,
which is body centered tetragonal in the case
of the ferrous alloys.\cite{cn:Matsui80,cn:Muto88,cn:Cui04}

For the above mentioned alloys, the relevant electron concentration
range is achieved at compositions which may allow the
formation of ordered stoichiometric compounds.
However, Fe$_3$Pt is the only alloy where the ordered phase is
reproducibly realized in experiment.
Nevertheless, measurements are predominately made for
slightly off-stoichiometric ordered
Fe$_{72}$Pt$_{28}$ which does not transform martensitically, even if
order is not fully complete.
Therefore, most data
are available only for disordered or partially ordered alloys.
Despite this fact, we will restrict our study to
stoichiometric, ordered compounds. These are much easier to handle
from the technical point of view and allow deeper insight into
electronic and vibrational properties with respect to their physical origin,
because their simulation cells are sufficiently small while
side-effects as statistical broadening can be largely avoided.

Another important feature of these alloys
is the marked softening of the
transversal acoustic phonons in [110] direction, which has
subsequently been related to any of the three above mentioned
anomalies in the past.\cite{cn:Tajima76,cn:Sato82,cn:Noda83,cn:Noda88,cn:Schwoerer96a,cn:Schwoerer96b,cn:Kaestner99a,cn:Kaestner99b,cn:Maliszewski99}
Our central aim is thus to investigate the influence of
electron-phonon coupling in Fe$_3$Ni, Fe$_3$Pd  and  Fe$_3$Pt
as one of the important mechanisms for structural
transformations of Fe-based binary alloys.  
This complements the experimental information by providing a link
between electronic properties and structural distortions, which will help to
clarify the microscopic origin of these instabilities.
We accomplished this aim by comparing
features of the phonon dispersion with the electronic structure of
these alloys. A similar approach has been used in the past to explore
the nature of electron-phonon coupling in conventional and magnetic
shape memory alloys.\cite{cn:Lee02,cn:Bungaro03,cn:Zayak03,cn:Hickel08,cn:Gruner09Review}

%%%%%%%%%%%%%%%%%%%%%%%%%%%%%%%%%%%%%%%%%%%%%%%%%%%%%%%%%%%%%%%%%%%%%
\begin{figure}[tb] 
  \begin{center}  
\includegraphics*[width=6cm]{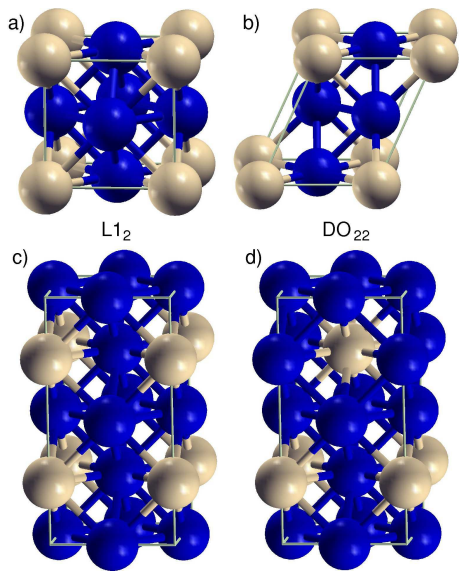}
  \end{center}
  \caption{Primitive cells of fcc L1$_2$ (a) and D0$_{22}$ (b) ordered
  alloys of $A_3B$ stoichiometry as used in our calculations.
  The lower two images show both structures represented in the same
  supercell which consists of two primitive cells stacked along the
  $c$-axis in the case  of L2$_1$ (c). Exchanging the $A$ and $B$
  components (marked by dark blue and bright orange spheres) in one of
  the $a$-$b$-planes yields the D0$_{22}$ ordering (d).
  Note, that the symmetry of the D0$_{22}$ structure is tetragonal.
  After reduction of the $c/a$-ratio by a factor of
  $\sqrt{1/2}$, the cell transforms to the
  bcc-type D0$_3$ structure with cubic symmetry.
}
\label{fig:cell}
\end{figure}
%%%%%%%%%%%%%%%%%%%%%%%%%%%%%%%%%%%%%%%%%%%%%%%%%%%%%%%%%%%%%%%%%%%%%

During our preoccupation with this topic, we noticed that investigations
providing a structural and dynamical view of the field on the basis of
first principles calculations are sparse and performed with
varying methodology,
hindering a comparison of all three isoelectronic
alloys on equal footing.
We therefore decided to provide first a systematic survey
of structural properties and lattice dynamics of
ordered Fe-rich alloys with elements of the Ni-group,
thereby filling the
above mentioned gap in existing literature.
The experimentally observed order in
Fe$_3$Pt is of L1$_2$ type, see Fig.\ \ref{fig:cell}. In our
calculations, we will alternatively consider D0$_{22}$ order.
For lattices with bcc coordination D0$_{22}$ becomes equivalent
with D0$_3$ order, which is related to
the L2$_1$ structure of ternary Heusler alloys.

Fe-rich Fe-Pd and Fe-Pt alloys close to the stoichiometric composition
are reported to be collinear ferromagnets at low temperatures; we will therefore
solely concentrate on the collinear ferromagnetic
case. Excited non-collinear,
ferri- and antiferromagnetic
spin structures, which have been discussed with
respect to the Invar effect,\cite{cn:Uhl94,cn:Schilfgaarde99,cn:Abrikosov07}
would certainly be beneficial for a thorough understanding
of the interdependence of lattice dynamics and finite temperature anomalies
as MSM and Invar effect.
These, however, are beyond the scope of the current work
and are thus left open for future investigation.

\section{Numerical details}
%%%%%%%%%%%%%%%%%%%%%%%%%%%%%%%%%%%%%%%%%%%%%%%%%%%%%%%%%%%%%%%%%%%%%
\begin{figure}[tb] % Fig. 2a and 2b
  \begin{center}  
\includegraphics*[width=8cm]{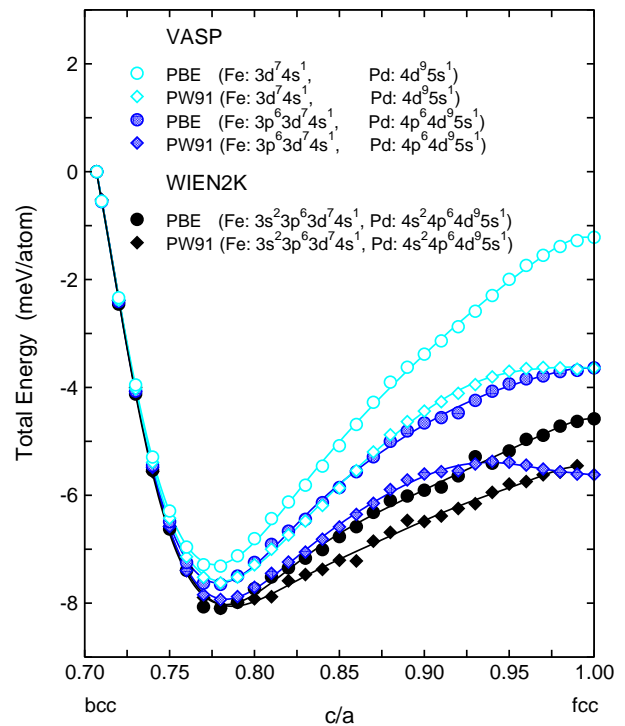}\\
  \end{center}
  \caption{The energy profile along the Bain path
    for L1$_2$ Fe$_3$Pd obtained with {\sc VASP} and {\sc Wien2k}
    for PBE and PW91
    exchange-correlation potentials. The calculations were carried out
    at a fixed atomic volume of 13.08\,\AA$^3$, corresponding to an
    fcc lattice constant of $a$$\,=\,$3.74\,\AA.
    For better comparison, all results are given relative to the
    energy of the bcc state ($c/a$$\,=\,$$\sqrt{1/2}$).
    The labeling fcc and bcc refers to coordination of the atoms
    regardless of the species.
    With the {\sc VASP} code,
    pseudopotentials with different numbers of
    semi-core electrons were tested;
    the reference calculations ({\sc Wien2k}) 
    account for Fe\,3$s$ and Pd\,4$s$ as semi-core electrons.
    Most curves coincide along the full Bain path within
    2\,meV/atom.}
\label{fig:VASPWIEN}
\end{figure}
%%%%%%%%%%%%%%%%%%%%%%%%%%%%%%%%%%%%%%%%%%%%%%%%%%%%%%%%%%%%%%%%%%%%%
Our investigation of the structural and dynamic properties has been
performed within the framework of density functional theory
(DFT).\cite{cn:Hohenberg64}
The majority of our results has been obtained using the
Vienna ab initio Simulation package ({\sc VASP}). \cite{cn:VASP1,cn:VASP2}
This code concentrates -- as pseudopotential methods generally do --
%
% Referee 1
%
for efficiency reasons
on the valence electrons for the description of the
electronic structure.
The {\sc VASP} code provides, nevertheless, an excellent compromise
between speed and accuracy as
the interaction with core electrons is taken into account
within the projector
augmented wave approach (PAW).\cite{cn:Bloechl94a} This yields results
close to what is usually expected from other all electron methods.
As we are dealing with Fe-rich compositions,
the use of the generalized gradient approximation (GGA) is mandatory
to obtain a correct description of the structural
ground state properties.
For structural energies within small (four atom) cells 
corresponding to Fig.\ \ref{fig:cell} (a) and (b) we use a $k$-mesh
containing at least 16384 $k$-points$\,\times\,$atoms
in the full Brillouin zone.
For the calculation of
the density of states (DOS) 
the $k$-point density was increased to values of 131072
$k$-points$\,\times\,$atoms and
340736 $k$-points$\,\times\,$atoms for obtaining the Fermi surface in
the 4 atom cell.
The Brillouin zone integration
was carried out using the tetrahedron method with Bl\"ochl
corrections.\cite{cn:Bloechl94b}
Electronic self-consistency was assumed below
a threshold difference of $1\,\mu$eV between two iteration.
The {\sc VASP} code is supplied with an extensive, well tested
potential library which
contains more than one pre-generated pseudopotential
per element. These differ by number of electrons explicitly
treated as valence or semi-core states or are designed for
different exchange correlation functionals. In our
case, the GGA formulations of Perdew and Wang (PW91) in connection with
the spin-interpolation formula of Vosko, Wilk and Nusair
as well as Perdew Burke and Ernzerhof (PBE) have been
used.\cite{cn:Perdew91,cn:Vosko80,cn:Perdew96,cn:Perdew96a}
The required cutoff of the plane wave basis depends on the exchange correlation
potential and was chosen as E$_{\rm cut}=335\,$eV (or larger) for PW91 and PBE
together with the minimal basis explicitly describing the
$3d^74s^1$ for Fe and $4d^95s^1$ for Pd (and correspondingly for Ni and
Pt). In calculations explicitely considering the semi-core $p$-electrons,
E$_{\rm cut}$ was set to $367\,$eV for PBE and $438\,$eV for the
PW91 pseudopotentials.  
A scalar relativistic formulation of the
Hamiltonian was used throughout.\cite{NOTE:RELATIVISTIC}

\subsection{Comparison between VASP and Wien2k}
Although the above mentioned technical differences with respect to the
choice of the exchange correlation functional and the basis size
usually do not lead to qualitative changes of the results,
we must be particularly careful in our case, because energy landscapes
in the vicinity of a structural transformation can be essentially flat.
This requires a higher accuracy, i.\,e., a better energy resolution.
Therefore, we decided to perform in a first step a thorough
comparison of the available potentials.
As a benchmark, we evaluated the energy
as a function of the tetragonal distortion $c/a$ along the
Bain path\cite{cn:Bain24}
for L1$_2$ ordered Fe$_3$Pd, describing a structural deformation from
a face centered cubic (fcc) lattice at $c/a=1$ to
a body centered cubic (bcc) lattice at $c/a=\sqrt{1/2}$, as shown in
Fig.\ \ref{fig:VASPWIEN}.
The results were compared
to high precision calculations using the full potential linearized
augmented plane wave method (FLAPW) as implemented within
{\sc Wien2k} code,\cite{cn:Wien}
which is widely counted among the most accurate
DFT codes for solid state problems. Here, the calculations included $3s$ and
$3p$ of iron and $4s$ and $4p$ of palladium as semi-core.
The energetic cut-off was chosen as $R_{\rm MT}$$\,\times\,$$K_{\rm
  max}$$\,=\,$9.0.
Muffin-tin radii of 2.24 a.u.\ were used for all atoms and
angular momenta were taken into account up to $l_{\rm
  max}$$\,=\,$10. The $k$-mesh comprised again
32000\,$k$-points$\,\times\,$atoms in
the irreducible Brillouin zone.

The obtained energy variation along the Bain path is with
$\approx 8\,$meV/atom very small. However, except for the PBE potentials
without semi-core $p$ electrons, all curves coincide within an energy
interval of 2$\ldots$3\,meV/atom (23$\,\ldots\,$35\,K on
a temperature scale), which
can be taken as a measure for the methodological resolution in
calculating structural energy differences along the Bain path.
While the PBE potentials seem to systematically overestimate the
energy variation (slightly), the agreement between 
both PW91 potentials (large and small basis)
and the full potential curves is very good. This finding agrees well with
a recent comparison of experimental and theoretical structural data
for equiatomic FePt alloys.\cite{cn:Zotov08}
Our benchmark demonstrates that the PW91 pseudopotentials with the small
basis (Fe $3p$ and Pd $4p$ as core electrons) is sufficient,
which is especially helpful for expensive calculations of large systems.

\subsection{Calculation of dynamical properties}
Two standard techniques are currently applied for the
investigation of lattice dynamics of crystals from
first principles: The linear response method and
the so-called direct method.\cite{cn:Kresse95,cn:Gonze97,cn:Parlinski97}
 
In the linear response method the dynamical matrix is
obtained from the modification of the electron density, via the inverse
of dielectric matrix describing the response of the valence electron
density to a periodic lattice perturbation. 
The dielectric matrix is then calculated from the
eigenfunctions and energy levels of the unperturbed
system.\cite{Resta1985}
Only linear effects, such as harmonic phonons, are accessible to this 
technique. The method has been applied with success to many 
alloys related to our present study.\cite{cn:Corso00,cn:Bungaro03}

On the other hand, the direct-method is a frozen-phonon type of
calculations based on a supercell calculation, which allows 
explicit account of any distortion of the atomic positions.
The phonon
frequencies are calculated from Hellmann-Feynman forces generated by the
small atomic displacements, one at a time. Hence using the information of
the crystal symmetry space group, the force constants are derived, the
dynamical matrix is built and diagonalized, and its eigenvalues arranged into
phonon dispersion relations. In this way, phonon frequencies at selected 
high-symmetry points of the Brillouin zone can be
calculated.\cite{cn:Parlinski05} The direct approach in conjunction with
the {\it ab}-initio method has previously been used intensively
by the authors to investigate
phonon dispersion relations in magnetic shape memory Heusler
compounds.\cite{cn:Zayak03,cn:Zayak05,cn:Entel08ESOMAT,cn:Gruner08EMRS}

The results presented within this work are relying on the direct approach
together with the {\sc VASP} code for the calculation of the
respective forces; the displacements being necessary to describe the
phonon dispersions were generated by the {\sc PHON} code
written by Dario Alf\`e{}.\cite{cn:Alfe09,cn:PHON}
This code was also used later
on to generate the dynamical matrix and the resulting phonon
dispersion relations. We used a supercell of 5$\times$5$\times$5
primitive cells, containing 500 atoms in total. For sufficient
accuracy of the forces a $k$-mesh of 4$\times$4$\times$4
points in the full Brillouin zone was employed in connection with
the Methfessel-Paxton finite
temperature integration scheme (smearing parameter
$\sigma$$\,=\,$0.1\,eV).
First-principles calculations of this size are computationally
very demanding and can,
so far, only be performed on world leading supercomputer installations.
The forces were calculated
for three independent displacements of $0.02\,$\AA{} in size each.
From these calculations
the phonon dispersions along the lines connecting the main
symmetry points and the vibrational density of states (VDOS) were
calculated. For the latter, a mesh of 61$\times$61$\times$61 points in
reciprocal $q$-space was used and additional Gaussian broadening with a
smearing parameter of $\sigma$$\,=\,$0.1\,THz was applied.
\medskip

We also carried out comparative calculations for
D0$_3$ Fe$_3$Ni and L1$_2$ Fe$_3$Pt using the linear response method
as implemented in the PWSCF package.\cite{NOTE:PWSCF}
To describe the interaction 
between ionic cores and valence electrons,
we used for Fe$_3$Ni ultra-soft
pseudo-potentials\cite{cn:Vanderbilt90} generated
using the exchange correlation functional of Perdew, Burke and Ernzerhof 
(PBE).
The pseudopotentials for Fe$_3$Pt were generated for the
exchange correlation of Perdew-Zunger within local density
approximation (LDA).\cite{NOTE:PSEUDO}
The technical parameters used here, were
a kinetic energy cutoff of 50 Ry, an energy cutoff for the 
augmentation charges of 600 Ry and 12$\times$12$\times$12 $k$-points in the full
Brillouin zone (Monkhorst-Pack). For 
the phonon dispersions, a mesh of 4$\times$4$\times$4 $q$-points in the
reciprocal space was used for the fcc basis, yielding 8 sets
with finite weight while for the simple cubic basis a 
2$\times$2$\times$2 $q$-points mesh was used yielding
4 sets of special $q$-vectors with finite weight.
Although the technical parameters are less tight,
the obtained dispersions agree well on a qualitative level
with the ones obtained within the direct approach
(except for discrepancies which can be partially related to the use
of LDA in the
latter case as the typical underestimation of the lattice constant as
well as an extension of the instability of the TA$_1$ branch to
the $\Gamma$-point, see also Ref.\ \onlinecite{cn:Adeagbo06})
and are therefore not explicitly presented in this manuscript.

%%%%%%%%%%%%%%%%%%%%%%%%%%%%%%%%%%%%%%%%%%%%%%%%%%%%%%%%%%%%%%%%%%%%%
%%%%%%%%%%%%%%%%%%%%%%%%%%%%%%%%%%%%%%%%%%%%%%%%%%%%%%%%%%%%%%%%%%%%%
%%%%%%%%%%%%%%%%%%%%%%%%%%%%%%%%%%%%%%%%%%%%%%%%%%%%%%%%%%%%%%%%%%%%%

\section{Comparison of structural properties}
%%%%%%%%%%%%%%%%%%%%%%%%%%%%%%%%%%%%%%%%%%%%%%%%%%%%%%%%%%%%%%%%%%%%%
\begin{figure} % Fig. 2a and 2b
  \begin{center}  
\includegraphics*{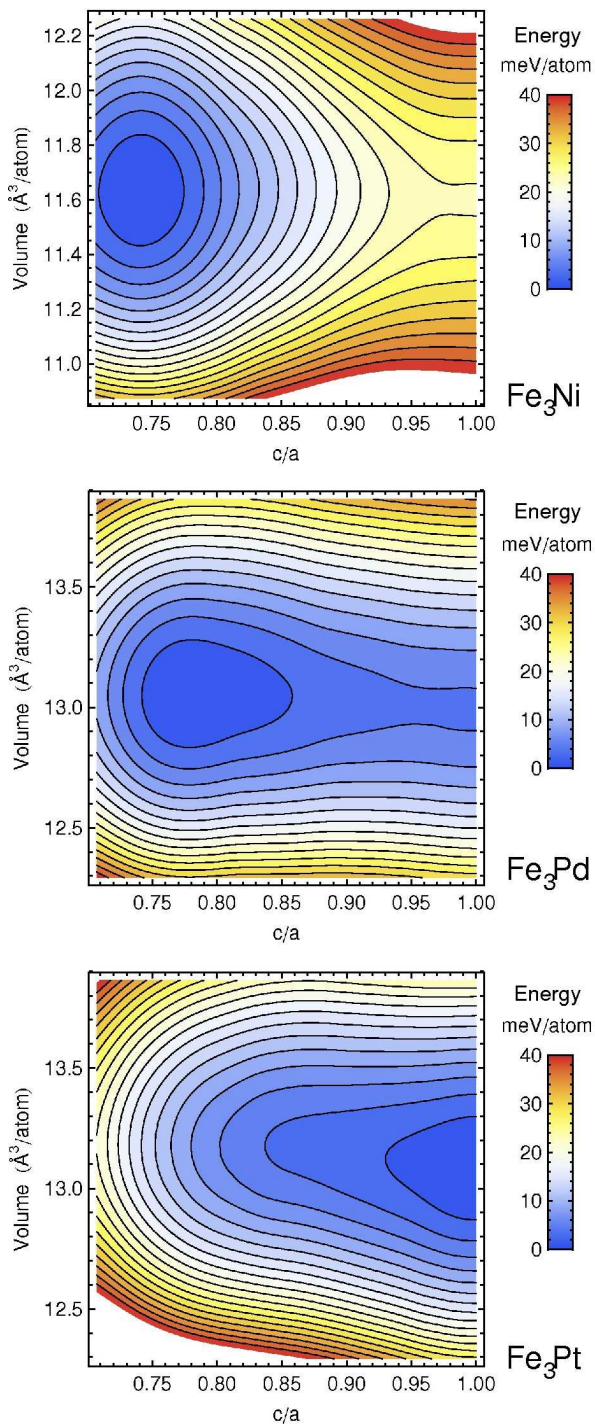}\\
  \end{center}
  \caption{(color online)
    Contour plots of the binding surfaces (energy as a function
    of atomic volume $v$ and tetragonal distortion $c/a$) of ferromagnetic
    L1$_2$ ordered Fe$_3$Ni (top),  Fe$_3$Pd (center) and Fe$_3$Pt (bottom).
    The energy distance between two contour lines is 2\,meV/atom.
    The global minima are located at 
    $c/a=0.74$ and $v=11.633\,$\AA$^3$/atom for Fe$_3$Ni,
    $c/a=0.78$ and $v=13.054\,$\AA$^3$/atom for Fe$_3$Pd,
    and $c/a=1.00$ and $v=13.082\,$\AA$^3$/atom for Fe$_3$Pt.}
\label{fig:contour}
\end{figure}
%%%%%%%%%%%%%%%%%%%%%%%%%%%%%%%%%%%%%%%%%%%%%%%%%%%%%%%%%%%%%%%%%%%%%
There are a large number of reports of first principle investigations
concerning Fe-rich alloys with elements of the platinum group
available in literature. These cover magnetic and magneto-elastic
properties,\cite{cn:Moruzzi89,cn:Moruzzi90,cn:Johnson90,cn:Entel93,cn:Schroeter95,cn:Khmelevskyi03,cn:Khmelevskyi04} electronic structure,\cite{cn:Podgorny91,cn:Major03,cn:Entel93,cn:Opahle09} order and disorder.\cite{cn:Taylor91,cn:Chen02,cn:Stern03}
Only few reports, however, are present describing the energetics of tetragonal
distortions in ordered alloys. These exist,
e.\,g., for Fe$_3$Pt using the LMTO-ASA (linear muffin-tin orbital
method within the atomic sphere approximation)
method
and for Fe$_3$Ni using different
approaches.\cite{cn:Lipinski99,cn:Entel93,cn:Hoffmann93,cn:Herper95}
Again, particularly the FLAPW calculations of
Ref.\ \onlinecite{cn:Herper95} agree very well with our results.
A systematic {\em ab-initio} comparison of all three alloys
with respect to structural properties
along the Bain path,
describing the transformation from fcc-type to bcc-type structure,
%
% Referee 1
%
however, is still missing
and therefore concisely presented in the subsequent section.

\subsection{Binding surfaces along the Bain path}
The total energy as a function of atomic
volume and tetragonal distortion for all three isoelectronic alloys
in L1$_2$ order is shown in Fig.\ \ref{fig:contour}.
As is easily seen from the contour plots, the optimum $c/a$ ratio
varies from close to bcc ($c/a=0.74$) for Fe$_3$Ni over bct
($c/a=0.78$) for Fe$_3$Pd to fcc ($c/a=1$) for Fe$_3$Pt.
This trend cannot simply be interpreted in terms
of the atomic volume as, especially for Fe$_3$Pd and Fe$_3$Pt, the
equilibrium volumes nearly coincide. This is in good agreement with
experimental measurements (see the collection of Okamoto,
Ref.\ \onlinecite{cn:Okamoto93}, for an extensive overview). Especially
the relation between lattice parameter and composition of  Fe$_3$Pd
exhibits a strong positive deviation from Vegards law.
The overall variation of the energy landscape along the Bain path is
exceptionally small for all three alloys. It is with 22\,meV/atom
largest for Fe$_3$Ni and smallest for  Fe$_3$Pd with only 4\,meV/atom
(minimum to saddle point).
The shift of the minima 
agrees well with the fact that in the
phase diagram of ordered Fe$_3$Pt
the stability range of the body centered structures at
$T$$\,=\,$$0\,$K ends at higher Fe-content compared to the other alloys
as well as with the trend in the martensitic transition
temperatures, which are reduced on variation of the second element
from Ni to Pt.
The maximum change in equilibrium volume associated with a full tetragonal
transformation along the Bain path is small,
less than one percent for all
three alloys. 
There is also a strong experimental indication for
an unusually flat energy surface.
Recently, it has been shown that thin disordered Fe$_{70}$Pd$_{30}$ films
can be grown epitaxially on different substrates
inducing different lattice constants $a$ in the
film plane.\cite{cn:Buschbeck09} The corresponding perpendicular lattice
constant, $c$, however, is free to adjust, thereby realizing $c/a$
ratios covering most of the Bain path. Such a strained epitaxial growth can
only be expected if the energy associated with the lattice strain
is sufficiently small along the Bain path.

%%%%%%%%%%%%%%%%%%%%%%%%%%%%%%%%%%%%%%%%%%%%%%%%%%%%%%%%%%%%%%%%%%%%%
\begin{figure*}[tb]
  \begin{center}  
\includegraphics*[width=\textwidth]{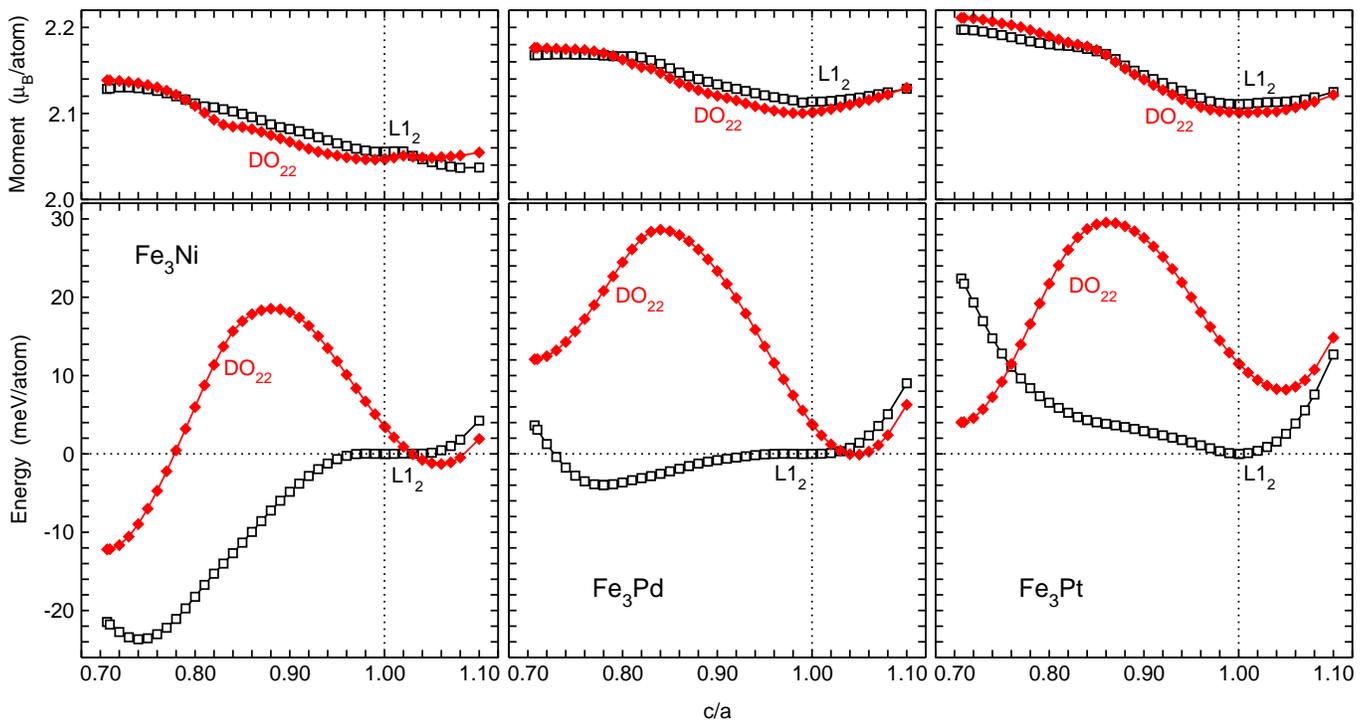}\\
  \end{center}
  \caption{(color online) Comparison of total energy and magnetic moment as a
   function of tetragonal
   distortion $c/a$ between L1$_2$-ordered and D0$_{22}$-ordered
   ferromagnetic Fe$_3$Ni (left),  Fe$_3$Pd (center) and Fe$_3$Pt (right). The
   calculations were performed at a constant atomic volume close to
   the respective equilibrium volume of the L1$_2$ structure.}
\label{fig:D022}
\end{figure*}
%%%%%%%%%%%%%%%%%%%%%%%%%%%%%%%%%%%%%%%%%%%%%%%%%%%%%%%%%%%%%%%%%%%%%

The  magnetic moment is largest on the bcc side, however, there
is only little variation of the spin moments along the Bain path,
which is of the order of
0.1$\,\mu_{\rm B}$/atom (cf., Fig.\ \ref{fig:D022}). The slight
variation of the magnitude between the alloys
might well be attributed to the different equilibrium lattice constants.

\subsection{L1$_2$ versus D0$_{22}$ order}
For fcc alloys of stoichiometry A$_3$B, the D0$_{22}$ structure may be
considered as another, possibly competing, realization of order.
As can be seen from  Fig.\ \ref{fig:cell} (d),
the D0$_{22}$ structure emerges from the
L1$_2$ structure by shifting one of the mixed planes
half way along the diagonal without shearing the other
planes.
At the c/a-ratio of the bcc lattice, the D0$_{22}$ structure
turns into the highly symmetric D0$_3$ structure, which is equivalent
to the Heusler-type L$2_1$ structure for a binary composition.
For Fe$_3$Ni and  Fe$_3$Pd, where order is not substantiated
in experiment, the D0$_{22}$ becomes favored at $c/a$ ratios slightly
above one, as demonstrated in the lower panels of Fig.\ \ref{fig:D022}.
%
% Referee 2
%
For Fe$_3$Pt the D0$_{22}$ structure becomes favored at the bcc
end. Only for this alloy, the energy of D0$_{22}$ becomes competitive
with L$2_1$, taking into account the complete Bain path.
The formation of a D0$_3$ phase in ordered near-stoichiometric
Fe$_3$Pt might further be hindered by the slow kinetics of the
necessary diffusive transformation at comparatively low temperatures.
However,  at the
fcc side, the small energetic distance between L1$_2$ and D0$_{22}$
order indicates a low penalty and thus a comparatively high
probability for anti-phase boundaries in the ordered phases.

The magnetic response of both order types to a tetragonal distortion is identical.
An important difference between  D0$_{22}$ and  L$2_1$ order is, however,
that along the Bain path a considerable energy barrier exists
for  D0$_{22}$ but not for  L$2_1$, which would slow down
the transformation kinetics, if D0$_{22}$ could be stabilized.

\section{Lattice dynamics and soft phonons}
%%%%%%%%%%%%%%%%%%%%%%%%%%%%%%%%%%%%%%%%%%%%%%%%%%%%%%%%%%%%%%%%%%%%%
\begin{figure*}[p]
  \begin{center}  
\includegraphics[width=0.85\textwidth]{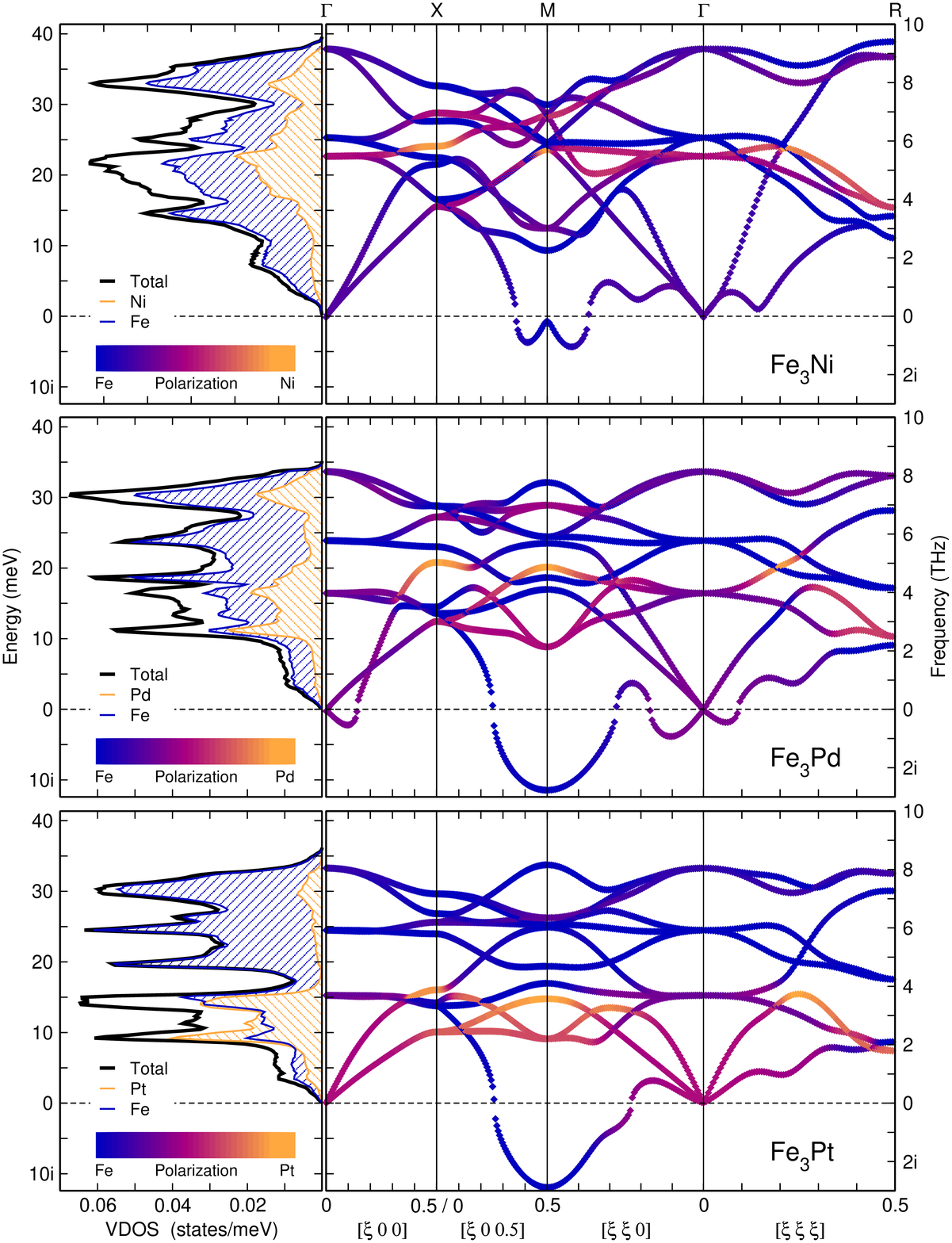}\\
  \end{center}
  \caption{(color online)
    Calculated phonon dispersions (right panels) of
    L1$_2$ ordered Fe$_3$Ni (top),  Fe$_3$Pd (center) and Fe$_3$Pt
    (bottom) along the main symmetry directions and the corresponding
    vibrational density of states (left panels). Imaginary frequencies
    are omitted from the density of states.
    The polarization of
    the phonon branches according to the atomic
    contributions to the respective eigenvectors is shown in the
    intensity (color) coding.
}
\label{fig:DISP-L12}
\end{figure*}
%%%%%%%%%%%%%%%%%%%%%%%%%%%%%%%%%%%%%%%%%%%%%%%%%%%%%%%%%%%%%%%%%%%%%
%
%%%%%%%%%%%%%%%%%%%%%%%%%%%%%%%%%%%%%%%%%%%%%%%%%%%%%%%%%%%%%%%%%%%%%
\begin{figure*}[tb]
  \begin{center}  
\includegraphics[width=0.9\textwidth]{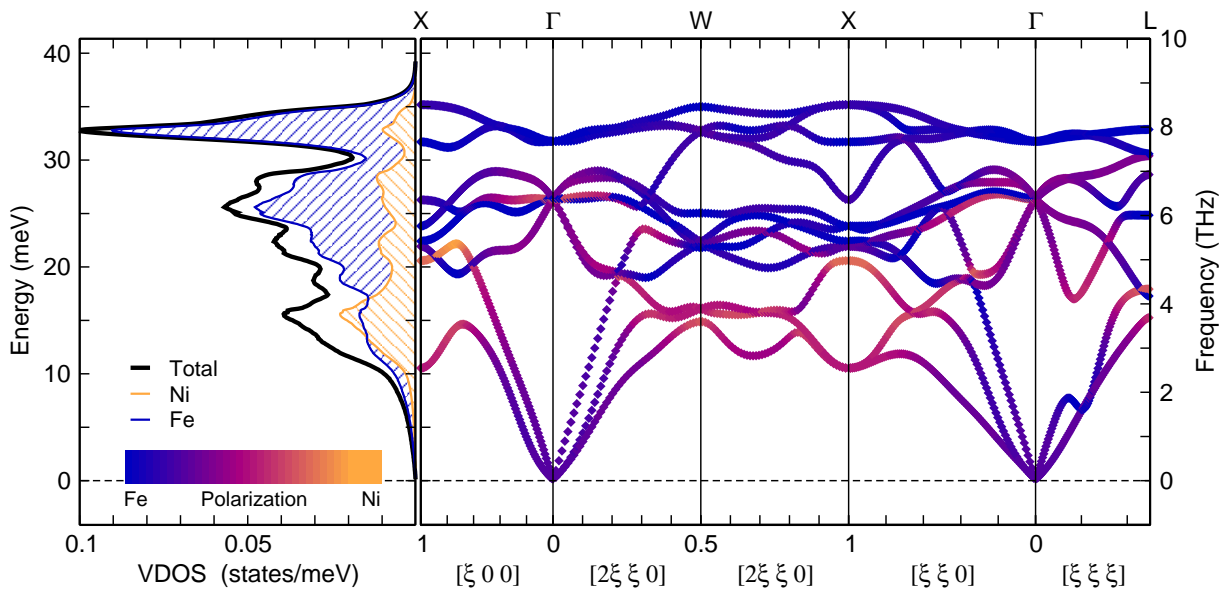}\\
  \end{center}
  \caption{(color online) Calculated phonon dispersions (right panel) of
    D0$_3$ ordered Fe$_3$Ni along the main symmetry directions
    and the corresponding
    vibrational density of states (left panel). Again, 
    the polarization of
    the phonon branches according to the atomic
    contributions to the respective eigenvectors is shown in the
    intensity (color) coding.
}
\label{fig:DISP-D03}
\end{figure*}
%%%%%%%%%%%%%%%%%%%%%%%%%%%%%%%%%%%%%%%%%%%%%%%%%%%%%%%%%%%%%%%%%%%%%
%
Important information about the stability of structures
and potential relaxation mechanisms
can be obtained from the phonon dispersion relations.
Accordingly,
extensive discussions have been taking place
concerning martensitic  precursor effects
in the lattice dynamics of the austenitic phase in martensitically
transforming alloys, which might yield, e.\,g., additional information on the
transformation mechanisms.
For the systems under consideration,
experimentally determined phonon dispersion relations 
obtained by  inelastic neutron scattering
are available for disordered
Fe$_{70}$Ni$_{30}$ and Fe$_{65}$Ni$_{35}$,\cite{cn:Hallman69,cn:Maliszewski99}
disordered Fe$_{72}$Pd$_{28}$,\cite{cn:Sato82} and
disordered as well as ordered Fe$_3$Pt.\cite{cn:Tajima76,cn:Noda83,cn:Noda88,cn:Schwoerer96a,cn:Schwoerer96b,cn:Kaestner99a,cn:Kaestner99b}
Theoretical calculations of the full phonon dispersion, on the other hand,
have been reported so far mainly for ordered
Fe$_3$Ni,\cite{cn:Herper95,cn:Adeagbo06},
disordered Fe$_{72}$Pd$_{28}$,\cite{cn:Akgun97,cn:Xavier07}
and Fe-Pd alloys in the Pd-rich composition range.\cite{cn:Ghosh08,cn:Dutta09,cn:Ghosh09}
Several of the above mentioned investigations rely on empirical or
semi-empirical descriptions of the interatomic forces.
Astonishingly, similar calculations for the Fe-Pt case appear to be
missing.
The inability of the experiment to provide a complete comparison of
the ordered phases (which are easier to understand) together with the
incomplete record of theoretical studies motivates
our attempt to provide a comparative overview of
the phonon dispersion of all three ordered isoelectronic alloys
from first principles.

\subsection{Phonon dispersion relations}
The central feature of the phonon dispersions is the complete
softening of the transversal acoustic branch TA$_1$
around the M-point.
This is common for all three alloys,
but most pronounced in Fe$_3$Pt.
The occurrence of imaginary frequencies reflects the
the instability of the lattice which
may gain energy by static atomic displacements according to the
corresponding phonon mode. 
A strong anomalous softening of the TA$_1$ branch
is a widely discussed observation from neutron diffraction
experiments on ordered
Fe$_3$Pt.\cite{cn:Tajima76,cn:Noda88,cn:Kaestner99a,cn:Kaestner99b}
With decreasing temperature starting from the paramagnetic phase, the 
TA$_1$ mode becomes increasingly softer
at the M-point, finally reaching phonon energies as low as 3\,meV for
$T=12$\,K. 
The microscopic nature of the corresponding displacements
will be discussed in the following section.
An incomplete softening of the TA$_1$ branch with anomalous
temperature dependence
is also found for the disordered alloys. However, in these cases,
it is far less pronounced and the wave vector is shifted towards the
$\Gamma$-point.\cite{cn:Sato82,cn:Maliszewski99,cn:Kaestner99a}

For Fe$_3$Pd one of the transversal acoustic modes in
Fig.\ \ref{fig:DISP-L12}
starts with a
negative slope from the $\Gamma$-point, which indicates
that the system might immediately undergo a long-wavelength
distortion without activation barrier. This is not the case for 
Fe$_3$Ni, although here, the underlying structure is not the
groundstate, either.
% Ref. 2
In both cases,
the cubic L1$_2$ state describes only a saddle point. Thus
the initial slope of the acoustic branches
in the low energy and low wavevector regime
depends on small details of the binding surfaces and
might well be affected by the specific choice of the method
and technical setup,
as demonstrated in Fig.\ \ref{fig:VASPWIEN}.
The softening along the [111] direction is strongest for Fe$_3$Ni.
Both TA branches are degenerate, 
reaching frequencies as low as 1\,meV around
$\xi$$\,=\,$0.14. 
The corresponding eigenvectors are close to 
$(1,-1,0)$ and
$(1,1,-2)$ 
and practically uniform for all four atoms in the primitive cells.
This may correspond
to the primary [11$\overline{2}$] shear on (111) planes,
common for martensitic transformation mechanisms of Fe-based alloys
(e.\,g., see Refs.\ \onlinecite{cn:Wechsler60,cn:Nishiyama78,cn:Ahlers02}).

The phonon dispersion of Fe$_3$Ni as shown in the upper panel of
Fig.\ \ref{fig:DISP-L12} agrees well with previous {\em ab initio}
results for the same material obtained by the same method (but less
restrictive accuracy).\cite{cn:Adeagbo06}
Differences are mainly present
at the M-point, where the complete softening of the TA$_1$ mode is
less pronounced in the present study and slightly shifted away from the
Brillouin zone boundary. In addition, the softening in [111] direction
is more pronounced in our study.
Previous semi-empirical calculations based on a tight-binding
scheme
on the other hand did only show comparable softening if the strength of the
electron phonon coupling was artificially enhanced.\cite{cn:Herper95}
This underlines the necessity to fully take into account the
coupling between electronic and lattice degrees of freedom
within a first-principles approach.
Still, the convergence with respect to the subdivision in
reciprocal space for the electronic structure as well as for the
lattice dynamics remains an important issue.

Another quantity which allows for a detailed
comparison with experiment is
the element resolved vibrational density of states (VDOS) which is
shown in the diagrams on the left of the dispersions.
For alloys with a large mass difference, we expect a clear separation
between the elemental contributions. This is indeed the case
for Fe$_3$Pt, where we find a pseudo-gap at phonon energies around
17\,meV. Above this energy, there is only little contribution from Pt
vibrations, while we find a hybridization of Pt and Fe contributions
in the energy range below.
In Fe$_3$Pd, the Pd states move to higher energies,
filling up the pseudo-gap and there is increased hybridization of the
partial contributions at the optical bands in the uppermost part.
This trend consequently continues for
Fe$_3$Ni, moving the central peak of the partial VDOS of Ni up to an
energy of 22.5\,meV. In turn, the distinct low energy peak appearing
around 10\,meV for Fe$_3$Pt and Fe$_3$Pd has disappeared. 
In all three cases, there exists a significant contribution
of Fe-states in the low energy region ($\leq\,8\,$meV).
The main peaks of the calculated VDOS agree well with
experimental measurements of disordered Fe$_{65}$Ni$_{35}$ of Delaire
and coworkers\cite{cn:Delaire05} as well as on ordered Fe$_3$Pt by Wiele
{\em et al.}\cite{cn:Wiele99,cn:WieleThesis} and even recent measurements of the
Fe-partial contributions of ordered  Fe$_3$Pt nanoparticles by
Roldan-Cuenya {\em et al.}\cite{cn:Roldan09} using the 
M\"o\ss{}bauer approach. The pseudo-gap, however, is not showing up in
experimental VDOS as these generally experience a
stronger broadening, which may be attributed to finite temperature of
the measurements, resolution of the measurement devices and -- not at
last -- the incomplete ordering usually present in Fe$_3$Pt samples.

The disappearance of the sharp energy peak in the VDOS around 10\,meV
points out a qualitative difference between Fe$_3$Ni and
the other two alloys. In the latter case, these states are connected
with vibrational modes at the M and the R-point
which are dominated by motions of the Pt or, respectively, Pd atoms as
can be seen from the atomic polarization of the phonon dispersions
shown in the intensity (color) coding in Fig.\ \ref{fig:DISP-L12}. In
Fe$_3$Ni, however, the low lying modes in the vicinity of the M-point
mainly involve Fe-atoms, which makes the Fe-subsystem particularly soft
and susceptible to element specific distortions, e.g., due to
magnetoelastic coupling. In contrast, this feature completely disappears
in the body centered cubic phase of Fe$_3$Ni, which we represent in
our calculations for simplicity by the D0$_3$ ordered structure
(cf.\ Fig.\ \ref{fig:DISP-D03}), as this also possesses cubic symmetry.
Here, again, we find a rather clear separation between the Fe and the
Ni-contributions which now predominately
occupy the low lying branches at the Brillouin zone boundary.
In this sense, the situation bears similarity to the inversion of the
Ni and Ga modes which was reported for the magnetic shape memory alloy
Ni$_2$MnGa.\cite{cn:Zayak05,cn:Zayak06,cn:Entel06Review}

\subsection{Identification of the soft mode}
%%%%%%%%%%%%%%%%%%%%%%%%%%%%%%%%%%%%%%%%%%%%%%%%%%%%%%%%%%%%%%%%%%%%%
\begin{figure} % Fig. 1a and b
  \begin{center}  
\includegraphics*[width=8cm]{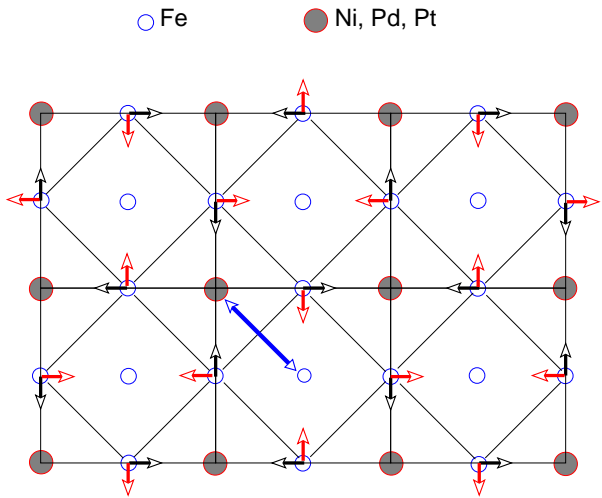}
  \end{center}
  \caption{(color online)
Schematic representation of the two M-point soft modes in the phonon
dispersions of fcc Fe$_3$Ni, Fe$_3$Pd and Fe$_3$Pt
shown in Fig.~\ref{fig:DISP-L12}.
The picture shows a projection of the repeated L1$_2$ unit cell
along the [0\,0\,1] direction. The brighter (red) arrows correspond to the M2 mode,
  while the black arrows show vibrations of the M4 mode. The main
  feature here is that both vibrations are identical, but orthogonal
  to each other.
  Both
  vibrations match with a translation along one half of the the face diagonal
  ([110] direction) as
  shown by the thick (blue) arrow, whereby the two vibrations remain
  the same but shifted in phase by  $\pi$.} 
\label{fig:pattern}
\end{figure}
%%%%%%%%%%%%%%%%%%%%%%%%%%%%%%%%%%%%%%%%%%%%%%%%%%%%%%%%%%%%%%%%%%%%%
As laid out in the previous paragraph,
the two lowest frequencies in the dispersion of L1$_2$
Fe$_3$Ni at the M-point involve only motions of the Fe species.
Since the associated wave-vector corresponds to the Brillouin zone boundary,
the amplitudes are opposite at either end of the primitive
cell. Therefore the term {\em antiferrodistortive transformation} has been
has been coined for the condensation of the unstable
phonon mode.\cite{cn:Kaestner99a}
The displacements can be inferred from the corresponding
eigenvectors and are sketched in Fig.\ \ref{fig:pattern}.
They belong to motions of the Fe-atoms either in direction of one of
the two nearest neighbor Ni pairs, or perpendicular to this direction.  
If one describes the primitive cell as an Fe-octahedron encaged by
Ni-atoms at the corner positions of the cube, the mode can be described
in the first case as a rotational motion of the perimeter atoms
in the (001) Fe-plane,
while the top and bottom atoms in the mixed plane
remain in place.
This kind of rotational mode is well known from
perovskite structures, where oxygen octahedra form such tilted
structures.\cite{cn:Ghosez99,cn:Parlinski00b}
In the second case, it is rather described as
a kind of breathing motion of the perimeter atoms in the
Fe-plane, where one half of the atoms move inwards and the other half
move outwards leading to an orthorhombic distortion of the Fe-octahedron.
Keeping the terminology of Noda and Endoh as well as K\"astner and coworkers,
we will call these modes M4 and M2, respectively.
%
% Referee 2
%
Displacements according to both phonon modes lower the symmetry of the
crystal and reduce the space group of the cubic L$1_2$ phase,
Pm$\overline{3}$m (number 221), to P4/mbm (number 127).
Since the latter is already
tetragonal, a variation of $c/a$ does not imply
another change of symmetry,
which corresponds well to the %above mentioned
experimental reports of a low temperature fct phase.
%
%%%%%%%%%%%%%%%%%%%%%%%%%%%%%%%%%%%%%%%%%%%%%%%%%%%%%%%%%%%%%%%%%%%%%
\begin{figure}[tb]
  \begin{center}  
\includegraphics*[width=8cm]{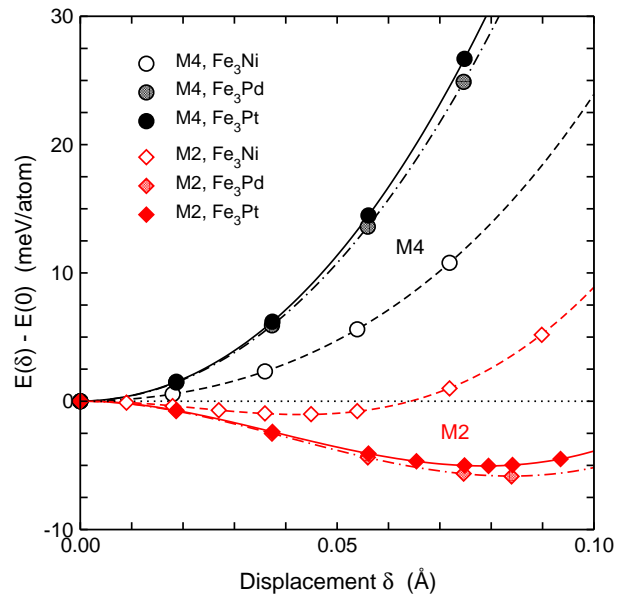}\\
  \end{center}
  \caption{(color online) Total energy as a function of the
    displacement amplitude according to the 
    two modes (M2, diamonds, and M4, circles) which may be
    made responsible for the
    for the (complete) softening of the acoustical phonons at the
    M-point (cf.\ Fig.\ \protect\ref{fig:pattern}).
    Results are shown for Fe$_3$Ni,  Fe$_3$Pd and
    Fe$_3$Pt. The lines are polynomials fitted to the calculated data
    points.
    For all three alloys, displacements according to the 
    M2-mode lead to a decrease in energy.}
\label{fig:M2M4}
\end{figure}
%%%%%%%%%%%%%%%%%%%%%%%%%%%%%%%%%%%%%%%%%%%%%%%%%%%%%%%%%%%%%%%%%%%%%

As mentioned above, the nature of the distortions
leading to the softening of the phonon modes can be inferred in
principle from the
eigenvectors of the dynamical matrix which can be constructed,
e.\,g., by a Born-von-Karmann fit to the experimental dispersions.
This indirect procedure often leaves room for ambiguities and
has lead to a controversy about the nature of the unstable mode in the
past.\cite{cn:Noda88,cn:Kaestner99a}
A more direct approach is to monitor the variation of the total energy
as a function of the displacement according to the (frozen) phonon modes. 
This can be realized in a straight forward manner within a density
functional theory approach. We construct 2$\times$2$\times$1
supercells by doubling the primitive L1$_2$ cell in the plane of the
atomic motion and apply the corresponding
displacements $\delta$ as sketched in Fig.\ \ref{fig:pattern}.
The result, shown in
Fig.\ \ref{fig:M2M4}, unambigously proves that for all three alloys
the M2 mode leads to a lower energy even for very small
$\delta$, explaining the imaginary frequencies at the M-point.
In Fe$_3$Pt and Fe$_3$Pd, the energy gain can reach values of about
5\,meV/atom for displacements of about 0.08 to 0.09\,\AA{}. In Fe$_3$Ni this
energy gain is significantly smaller and the minimum energy is obtained for
smaller displacements.
On the other hand, the M4 mode is considerably softer
in Fe$_3$Ni than in the other two alloys. 
Here, the two lowest frequencies correspond to the two pure Fe modes,
while for the other two alloys the second eigenvalue has a marked
contribution of the heavier element.
In the latter case, the M4 mode is represented
by the 5th and the 6th eigenvalue with frequencies of 18.7\,meV and
19.4\,meV for Fe$_3$Pd and Fe$_3$Pt, respectively, as opposed to
9.3\,meV for Fe$_3$Ni.

In this respect, it may be important to notice that a
translation of the mixed (001) plane in [1$\overline{1}$0] direction,
which leads to an exchange of the Fe and Ni-group atoms, transforms
both modes, M2 and M4, in each other.
Applying this transformation
once every second unit cell transforms the L$1_2$ in
D0$_{22}$ order, as can be seen from Fig.\ \ref{fig:cell}.
D0$_{22}$ can thus also
be regarded as an L$1_2$ structure with maximum density of
anti-phase boundaries.
Thus atoms in the Fe-plane will feel a superposition
between both environtments and displacements according to
M2 and M4 modes become equivalent.
Therefore, we expect that such
distortions of the order will likely harden the
soft phonon at the M-point, especially for Fe$_3$Pd and Fe$_3$Pt.

%%%%%%%%%%%%%%%%%%%%%%%%%%%%%%%%%%%%%%%%%%%%%%%%%%%%%%%%%%%%%%%%%%%%%
\begin{figure*}[tb]
  \begin{center}  
\includegraphics*[width=\textwidth]{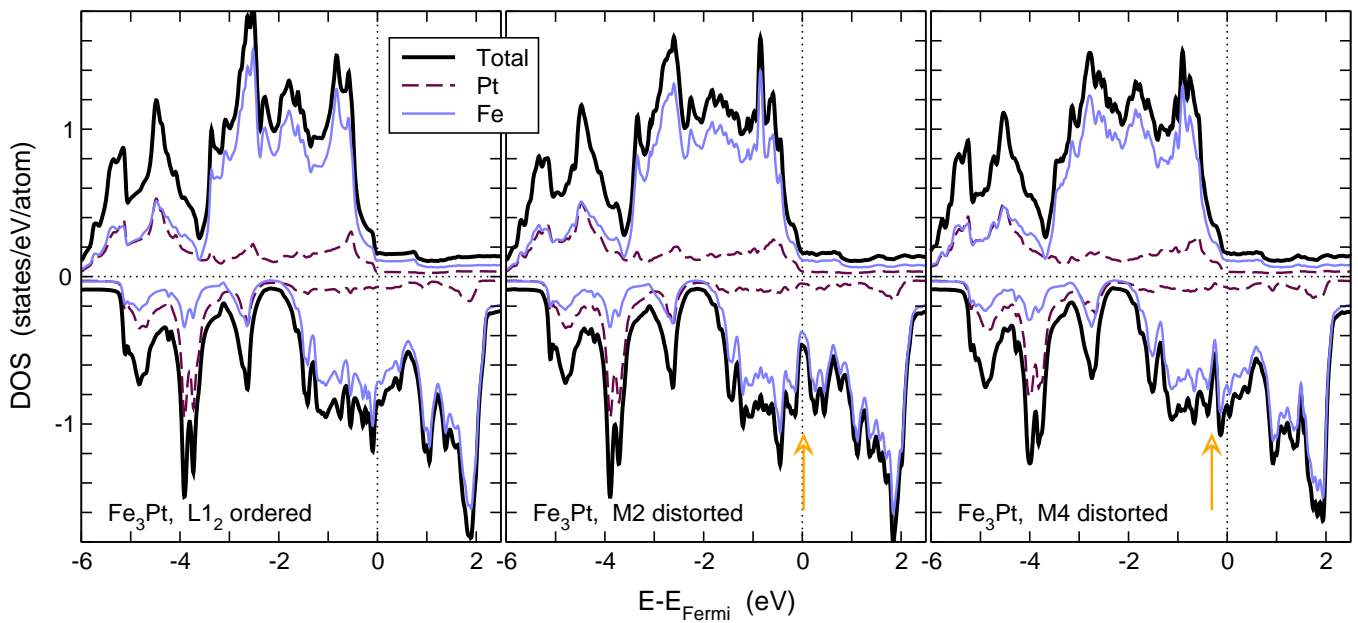}
  \end{center}
  \caption{(color online) Comparison of the total and partial, site resolved
    electronic density of states (DOS) of Fe$_3$Pt with perfect L$1_2$
    order (left) and M2 distorted  Fe$_3$Pt with an displacement amplitude
  $\delta=0.795\,$\AA{} (center) and M4 distorted  Fe$_3$Pt with
$\delta=0.748\,$\AA{} (right). The majority spin DOS are denoted by
    positive values, while negative values refer to the minority spin
    channel. 
 In the M2 case, a pseudo-gap opens
  at the Fermi level due to a redistribution of the Fe states.
 For the M4 distortion, a narrower gap opens up 0.25\,eV below the Fermi level,
 while the DOS at the Fermi level remains largely unchanged (pseudo-gaps
 denoted by arrows).
}
\label{fig:DOS}
\end{figure*}
%%%%%%%%%%%%%%%%%%%%%%%%%%%%%%%%%%%%%%%%%%%%%%%%%%%%%%%%%%%%%%%%%%%%%

\subsection{Electronic origin of the antiferro-distortive transformation}
While the total energy calculations supply us with the information that the
imaginary phonon mode is related to an
orthorhombic distortion of the Fe octahedra in the L1$_2$ unit cell,
we are still left with the task to find the origin of this instability.
A clue can be obtained from the comparison of the respective
electronic density of states, which
is shown paradigmatically
in Fig.\ \ref{fig:DOS} for the undistorted, perfect L1$_2$
structure of Fe$_3$Pt as well as for supercells with M2 and M4 distortions.
For all three cases, the majority spin density of states remains similar.
The L1$_2$ DOS is in excellent agreement with the previous extensive study of
Podg\`orny,\cite{cn:Podgorny91} apart from a small shift of the Fermi
level in both spin channels which can be attributed to the use of an
exchange correlation functional without gradient corrections in
Ref.\ \onlinecite{cn:Podgorny91}.
Fe$_3$Pt is at the onset of strong ferromagnetism,
thus the $d$-states of the majority
spin channel are filled and the respective DOS
is small at the Fermi energy, $E_{\rm F}$, leaving no room
for changes that could account for the structural distortions.
This, however, does not apply for the minority spin channel, where a
considerable number of $d$-states is encountered right at $E_{\rm F}$,
especially for the undistorted L1$_2$ structure. 
The vast majority of these states comes from the Fe atoms, while the
Pt DOS remains small, due to the nearly filled $d$-shell and only
moderate induced magnetic polarization of 0.36\,$\mu_{\rm B}$.

On the other hand, the condensation
of the M2-mode opens a deep pseudo-gap right at the Fermi-level
which is the indication
of an extensive redistribution of states.
In the spirit of a band-Jahn-Teller mechanism,
lifting the degeneracy of electronic states at the Fermi level
can lead to a net gain in band energy.
This has been previously proposed as the main electronic mechanism
leading to the tetragonal
distortion in disordered magnetic shape memory Fe-Pd alloys.\cite{cn:Opahle09}

Accordingly, a distortion corresponding to the M4 mode does not
decrease the number of states at $E_{\rm F}$. Nevertheless, it opens
another pseudo-gap 0.25\,eV below the Fermi level.
Although this gap is very narrow and should be expected to vanish
already at rather low
temperatures, it appears possible that a moderate decrease of the
valence electron concentration $e/a$ may bring this pseudo-gap to
the Fermi level and possibly causes a change the order of the modes
at the M-point which alters the corresponding ground state structure.

%%%%%%%%%%%%%%%%%%%%%%%%%%%%%%%%%%%%%%%%%%%%%%%%%%%%%%%%%%%%%%%%%%%%%
\begin{figure*}[tb]
  \begin{center}  
\includegraphics[width=0.5\textwidth]{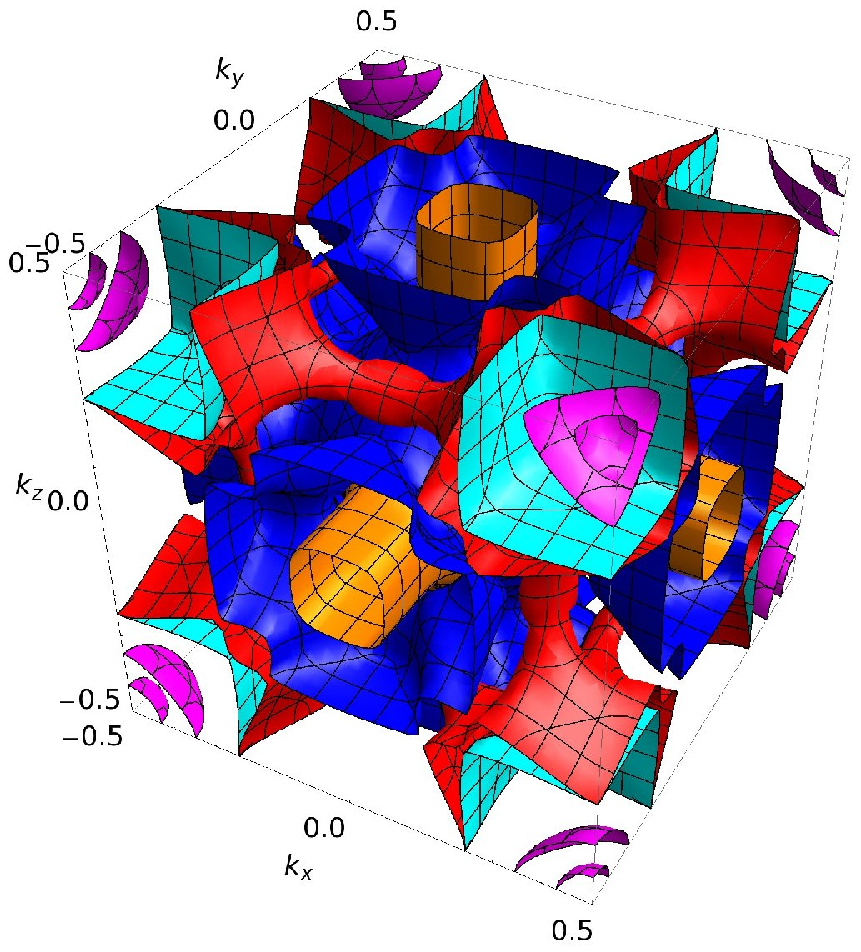}%
\hspace{-5mm}\includegraphics[width=0.5\textwidth]{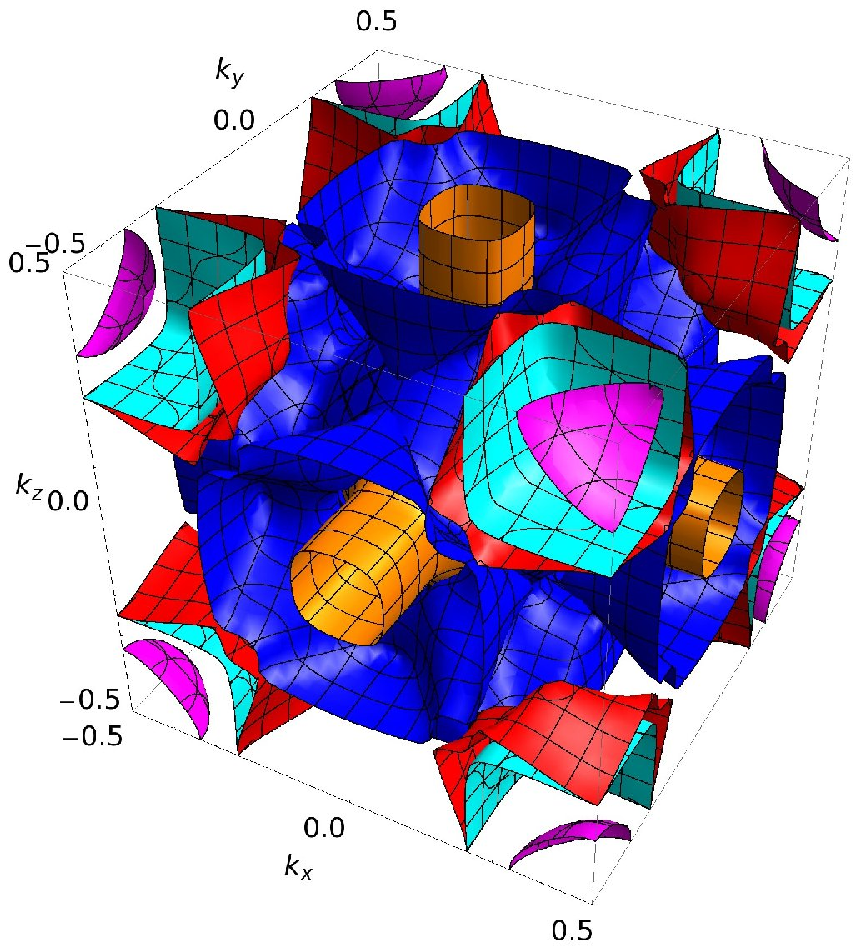}
  \end{center}
  \caption{Plots of the isoenergy-surface in reciprocal space of the
  minority spin electronic states the Fermi level (Fermi surface) of
  Fe$_3$Ni (left) and Fe$_3$Pt (right). The Fermi level is intersected
  by 6 bands (shown online in different colors). The features are largely
  similar and differ mainly only in size due to the varying atomic
  volumes. Both surfaces exhibit
  large nearly flat portions which may give rise to a Kohn anomaly.}
\label{fig:FS-both}
\end{figure*}
%%%%%%%%%%%%%%%%%%%%%%%%%%%%%%%%%%%%%%%%%%%%%%%%%%%%%%%%%%%%%%%%%%%%%
%
%%%%%%%%%%%%%%%%%%%%%%%%%%%%%%%%%%%%%%%%%%%%%%%%%%%%%%%%%%%%%%%%%%%%%
\begin{figure*}[tb]
  \begin{center}  
\includegraphics[width=\textwidth]{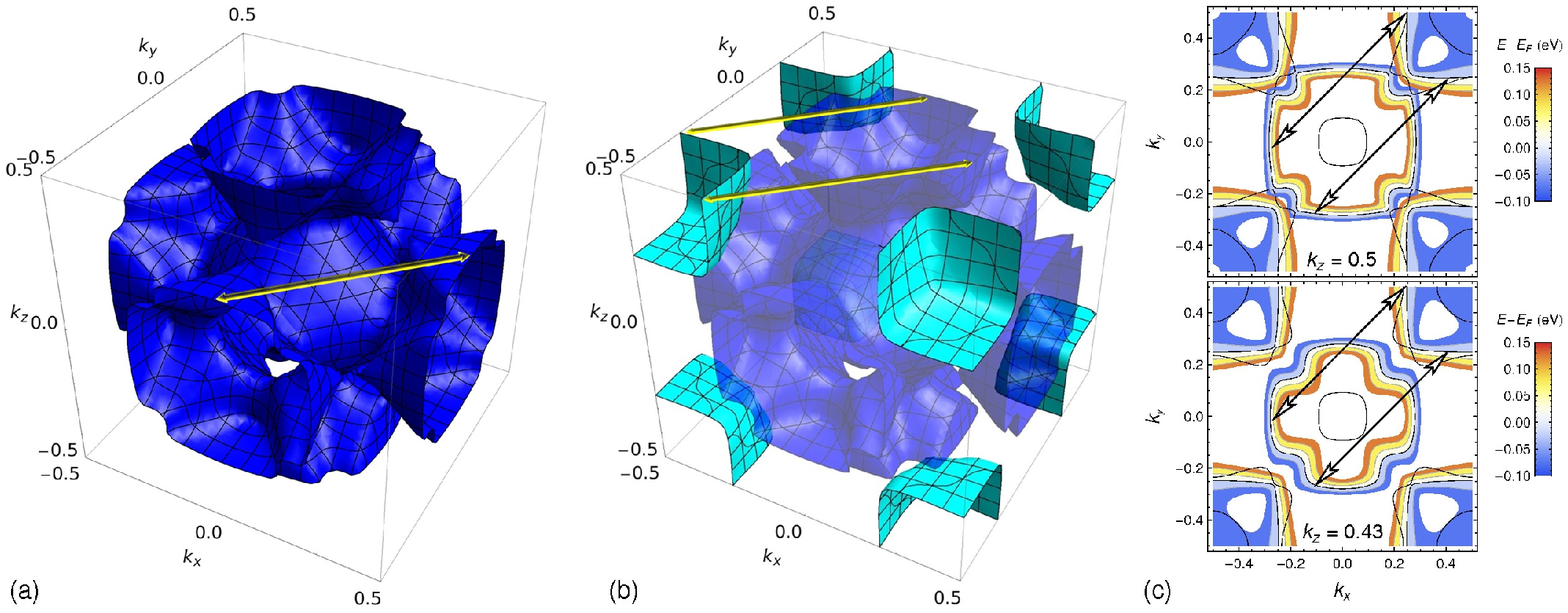}
  \end{center}
  \caption{Fermi surfaces of the 13th (a) and 15th (b)
   minority spin band of Fe$_3$Pt (with an additional semitransparent
   image of the 13th band in the latter case).
   These two bands exhibit
   flat parts which may be connected with a wave
   vector close to ($\frac{1}{2}$,$\frac{1}{2}$,0) as
   indicated by the arrows.
   The vector can either be applied tangentially between
   the flattened, horizontal walls of the cone-shaped extensions at the zone
   boundary entirely within the surface of the 13th band as shown in (a)
   or between the vertical walls on the left and the respective
   sections of the cube shaped surface centered around the R-point in (b).
   The eigenvalues in reciprocal space
   of the of the 13th and 15th bands
   along two horizontal planes ($k_z$$\,=\,$0.5 and
   $k_z$$\,=\,$0.43) are shown in (c) as contour plots.
   The intersection of the complete Fermi surface (all bands)
   with these planes are marked by black lines.
   The arrows denote again parallel
   vectors ($\frac{1}{2}$,$\frac{1}{2}$,0).
   The energies are given
   relative to the Fermi level; the difference between two
   contour lines is 50\,meV.
 }
\label{fig:FS-partial}
\end{figure*}
%%%%%%%%%%%%%%%%%%%%%%%%%%%%%%%%%%%%%%%%%%%%%%%%%%%%%%%%%%%%%%%%%%%%%
%

The appearance of a pseudo-gap at $E_{\rm F}$ implies a severe
reconstruction of the Fermi surface.
Such reconstructions in connection with anomalous softening of the
acoustic modes are a sign of strong electron-phonon coupling and
frequently related to a Kohn anomaly. This involves
a significant number of nesting states at $E_{\rm F}$
that can be connected with the same reciprocal
vector which describes the anomaly in the phonon dispersions.
Fermi-surface nesting and strong electron-phonon coupling in combination with 
transitions to modulated phases 
are commonly found in conventional and magnetic shape memory systems,
as Ni-Al, Ni-Ti and Ni-Mn-Ga.\cite{cn:Huang04,cn:Zhao93,cn:Lee02}
In fact,
the phonon softening in disordered Fe-Ni Invar
has been related previously to a Kohn anomaly in the nonmagnetic state
on the basis of first principles calculations of ordered
Fe$_3$Ni.\cite{cn:Hoffmann93}
Here, we consider spin-polarized minority spin Fermi surfaces, i.\,e., the
isosurfaces formed by the Kohn-Sham eigenvalues at the Fermi level
in reciprocal space.
These are depicted for Fe$_3$Ni and Fe$_3$Pt in
Fig.\ \ref{fig:FS-both} and we find good agreement of the shape
with previous results obtained using {\sc Wien2k}.\cite{cn:Wakoh02}

Six bands are crossing  $E_{\rm F}$ in
the minority spin channel making the Fermi surfaces quite complex objects.
Nevertheless, it is possible to relate nesting behavior to
two of these bands, the 13th and the 15th band which are shown
separately in Fig.\ \ref{fig:FS-partial}.
%%%%%
%
The 13th minority spin band represents an extended spherical
hole pocket around the $\Gamma$-point with hopper-shaped extrusions
reaching the Brilloin zone boundary.
Facilitated by the cubic symmetry, the flattened, horizontal walls of
these extrusions
can provide nesting with a vector of exactly
$(\frac{1}{2},\frac{1}{2},0)$, c.f., Fig.\ \ref{fig:FS-partial} (a).
These parts are connected by a tangential nesting vector,
which is insensitive to a specific value of $q$ and not expected to
produce a sharp contribution to the generalized
susceptibility. There is another possibility for nesting
between the Fermi surfaces of 13th and the 15th band.
The 15th band forms a cubic electron pocket around the R-point as depicted in
 Fig.\ \ref{fig:FS-partial} (b).
The walls of the cubes are nearly located on the same plane as the
flattened parts of the 13th band. This plane is parallel
to the boundary of the cubic Brillouin zone. Two of them exist in
each of the cubic directions, such that nesting can occur
between the two bands on the two different planes, respectively.
The correspondence in this case is not necessarily exactly
$(\frac{1}{2},\frac{1}{2},0)$
and in direction of the face diagonal,
but 
it can be expected to provide a pronounced peak in the generalized
susceptibility and thus a considerable contribution to electron-phonon coupling.
The cross sections in Fig.\ \ref{fig:FS-partial} (c), taken at two
different values of $k_z$, demonstrate that
this correspondence is not restricted to the zone boundary, but rather
affects a significant portion of the Fermi suface.
For Fe$_3$Ni the hopper-shaped extrusions of the 13th band
are smaller and the cubes larger than in the Fermi-surface of
Fe$_3$Pt, which results in a smaller distance between the parallel
planes and thus a smaller length of the $q$ vector. This coincides
nicely with the shift of the instability in the TA$_1$-branch
away from the M-point as well as the smaller magnitude of the imaginary
eigenvalue.

It should be mentioned that a thorough investigation
of the relation between Fermi surface nesting and transformative
processes requires a detailed statistics, e.\,g.,
an evaluation of the generalized susceptibility with respect to all
contributing reciprocal vectors as well as the calculation of the
electron-phonon coupling matrix elements. This is a formidable
task in itself and beyond the scope of this work. Nevertheless, in our view,
the purely graphical discussion above
gives already sufficient arguments that a Kohn
anomaly is responsible for the low temperature structural changes in
the ordered alloys.

\section{Conclusions}
Within the present investigation, we systematically
compared structural, electronic
and dynamic properties of ordered Fe-rich alloys with the elements of the
Ni-group.
The only one of these three alloy, which can be obtained experimentally
in the ordered state, Fe$_3$Pt,
is also the only one
which possesses an fcc ground-state structure. The dominant type of
ordering in all three alloys is L1$_2$, in accordance with
experimental observation. Competing D0$_{22}$ order was observed for
hypothetical Fe$_3$Ni and Fe$_3$Pd for fct structures with a
tetragonal distortion $c/a>1$
and for  Fe$_3$Pt with bcc coordination.
However, for ideal fcc ($c/a=1$), D0$_{22}$ order  is well within the
range of thermal energies, suggesting that anti-phase boundaries are
very likely to appear in the cubic L1$_2$ phases.

The phonon dispersions of the three alloys possess as a common feature a
complete softening of TA$_1$ branch in [110] direction around the
M-point, which is strongest for Fe$_3$Pt and Fe$_3$Pd as well as a partial
softening along the [111] direction. 
The first leads to the freezing of the respective phonon mode (M2)
which can be identified as an orthorhombic distortion of the Fe
octahedra in the primitive L1$_2$ cell
without involving the Ni-group elements.
The second anomaly becomes increasingly
pronounced with decreasing mass of the Ni-group element, as this leads
effectively to an increased Fe-contribution to the low energy
acoustical phonons.

It may be speculated whether the rather smeared out softening of the
TA$_1$ branch in [110] direction in the disordered alloys
can be interpreted in terms of a
corresponding distortion with similar electronic origin,
but on a larger, statistically modified period.
%
%% Referee 2
%
This owes to the fact that the Fe-octahedra
can be considered to be
statistically distributed within the bulk alloy and result
finally in a rather broadened and less pronounced anomaly.
These octahedra are characteristic for
the L1$_2$ structure, but also other locally Fe-enriched cluster
configurations, which are more or less susceptible
to deformations, may contribute.
Similarly, the gross electronic structure will
appear rather smeared out for the disordered crystal,
but distinct features may prevail locally within suitable
Fe-clusters and induce corresponding local distortions.
Indeed, characteristic relaxations of the atomic positions have been reported
in {\em ab initio} calculations for fcc FeNi-Invar alloys for 64-atom
quasi-random structures.\cite{cn:Liot09}
This also fits well with our observation that in Fe$_3$Ni not only the
M2 mode but also the orthogonal M4 mode
becomes rather soft at the M-point, which makes collective motions
in the Fe-(001) planes very cost effective.
Furthermore, own investigations
considering structural relaxations
of 108-atom Fe-Pd and Fe-Pt supercells with random distribution of
atoms, yield a considerable energy gain from a similar relaxation
process.\cite{cn:unpub}

The sum of these observations supports the view that the origin of
the anomalous features in the acoustical branches can be attributed
solely to the Fe species.
This is confirmed by the calculation of the electronic structure of the
L1$_2$ ordered and the distorted alloys. The condensation of the M2 mode
opens a pseudo-gap in the minority density of states right at the Fermi
level, reducing the density of states by nearly 50\,\%.
Again, only contributions from Fe states are involved.
The origin for this strong reconstruction of the Fermi surface can
be related to extended flat areas which nest with a reciprocal
vector of nearly $(\frac{1}{2},\frac{1}{2},0)$
which corresponds to the softening of the
TA$_1$ phonon mode at the M-point.

The subtle interdependence between electronic structure at the Fermi
level and the antiferro-distortive transformation confirms once more
that the valence electron density, which controls the overall filling
of the minority spin channel on a fine level, is one of the key factors
for the design of Fe-rich functional alloys.
On the basis of our investigation, we further conclude -- in accordance
with Ref.\ \onlinecite{cn:Opahle09} -- that
 the specific concentration of the Fe-species must be
considered as another important
quantity and, additionally, 
the mass of the alloying elements which modifies the lattice
dynamics related to the low lying acoustical modes.

%%%%%%%%%%%%%%%%%%%%%%%%%%
\section{Acknowledgments}
%%%%%%%%%%%%%%%%%%%%%%%%%%
The authors gratefully acknowledge helpful remarks and
discussions with S.\ F\"ahler (Dresden), W.\ Keune (Duisburg-Essen),
J.\ Neuhaus (M\"unchen), B. Sanyal (Uppsala) and
E.\ F.\ Wassermann (Duisburg-Essen).
Part of the calculations were carried out on the JUGENE supercomputer
of the John von Neumann Institute for Computing, Forschungszentrum J\"ulich.
We thank the staff of J\"ulich Supercomputing Center for their
continuous support. Financial support was granted
by the Deutsche Forschungsgemeinschaft in the frame of the priority
programme SPP 1239,
{\em Change of microstructure and shape of solid materials by external magnetic fields}.

%%%%%%%%%%%%%%

\begin{thebibliography}{99}
\bibitem{NOTE:RELATIVISTIC}
{% R. 1
It should be mentioned at this point,
that for heavy elements, as the $5d$ transition
metal Pt, a fully relativistic treatment would be superior in general.
In the present case, this also implies to go beyond
the frequently employed perturbative corrections, because
fully self-consistently obtained charge densities are needed
for the calculation of the inter-atomic forces, which are used
to determine the respective phonon dispersion relations.
Within the VASP code, spin-orbit terms are only available in
connection with a non-collinear representation of magnetic moments,
which significantly increases the
computational demands while slowing down convergence.
On the other hand, the benefit would be rather limited,
because the essential physics is governed by a redistribution
of Fe $3d$ states in the vicinity of the Fermi level,
as layed out in the further course of the paper.
For our purposes, we therefore restrict 
to the scalar relativistic approach,
allowing us to treat all three alloys consistently
on the same level of approximation at reasonable numerical cost.
}
\end{thebibliography}
\end{document}